\documentclass[aps,pra,twocolumn,showpacs]{revtex4}
\usepackage[dvips]{graphicx}
\usepackage{amsmath}
\usepackage{times}

\begin{document}

\title
{
Bose-Einstein-condensed gases in arbitrarily strong random potentials
}

\author{V.~I.~Yukalov,$^{1,2}$ E.~P.~Yukalova,$^3$ K.~V.~Krutitsky,$^1$ and R.~Graham$^1$}
\affiliation{
$^1$Fachbereich Physik der Universit\"at Duisburg-Essen, Campus Duisburg,
Lotharstr. 1, 47048 Duisburg, Germany\\
$^2$Bogolubov Laboratory of Theoretical Physics,
Joint Institute for Nuclear Research, Dubna 141980, Russia\\
$^3$Department of Computational Physics, Laboratory of Information Technologies,
Joint Institute for Nuclear Research, Dubna 141980, Russia
}

\date{\today}

\begin{abstract}
Bose-Einstein-condensed gases in external spatially random
potentials are considered in the frame of a stochastic self-consistent 
mean-field approach. This method permits the treatment of the system 
properties for the whole range of the interaction strength, from zero 
to infinity, as well as for arbitrarily strong disorder.
Besides a condensate and superfluid density, a glassy number density
due to a spatially inhomogeneous component of the condensate occurs.
For very weak interactions and sufficiently strong disorder, 
the superfluid fraction can become smaller than the condensate fraction, 
while at relatively strong interactions, the superfluid fraction is 
larger than the condensate fraction for any strength of disorder. The 
condensate and superfluid fractions, and the glassy fraction always coexist,
being together either nonzero or zero.
In the presence of disorder, the condensate fraction becomes a nonmonotonic
function of the interaction strength, displaying an antidepletion effect
caused by the competition between the stabilizing role of the atomic interaction
and the destabilizing role of the disorder.
With increasing disorder, the condensate and superfluid fractions
jump to zero at a critical value of the disorder parameter by a first-order 
phase transition.
\end{abstract}


\maketitle

\section{Introduction}

Physics of dilute Bose gases is usually studied (see Refs.~\cite{1,2,3,4,5,6,7})
for asymptotically weak interactions, when 
the Bogolubov approximation~\cite{8,9} is applicable. In the case of strong 
interactions, the most reliable techniques are purely numerical, such 
as Monte Carlo simulations~\cite{10,11,12,13,14,15,16,17}.
In the majority of experiments with trapped atoms,
interactions are rather small~\cite{3,4,5,6,7} corresponding to values of the
gas parameter much smaller than one. However, now
it has become possible to vary the interaction strength in a wide 
range by using the Feshbach resonance techniques (see Refs.~\cite{4,18}).
Thus, in experiments with $^{85}$Rb atoms~\cite{19,20,21}, the value 
of the gas parameter $0.8$ has been reached. Large values of the 
scattering length and, respectively, strong effective interactions 
can also be achieved in quasi-one-dimensional and quasi-two-dimensional 
configurations due to the geometric resonance~\cite{22,23} (see discussion 
in Refs.~\cite{4,7,24,25,26}).

Recently, based on the idea of representative statistical 
ensembles~\cite{27} as applied to Bose systems with broken gauge symmetry~\cite{28,29},
a self-consistent approach has been developed~\cite{30,31,32,33} for 
treating Bose-condensed systems with arbitrarily strong interactions. 
This approach was shown~\cite{33} to reproduce the weak-coupling expansions 
of Bogolubov~\cite{8,9} and Lee-Huang-Yang~\cite{34,35,36},
while simultaneously being 
in good agreement with numerical Monte Carlo simulations for strong 
interactions.

A fundamental feature of any uniform Bose system with arbitrarily strong 
interactions is the appearance, at low temperatures, of a Bose-Einstein 
condensate and, simultaneously, of superfluidity. In these systems, the 
condensate fraction, $n_0$, and the superfluid fraction, $n_s$, always 
coexist, both being nonzero below the condensation temperature. Though 
there is no simple relation between these fractions,
the superfluid fraction is always larger,
$n_s\geq n_0$.

When a uniform Bose system is subject to the action of an
external spatially random field, the relation between the condensate 
and superfluid fractions could change. Thus, Huang and Meng~\cite{37} 
considered a Bose-condensed system in a random external potential. 
They treated the case of asymptotically weak interactions and 
asymptotically weak disorder. Assuming that their results could be 
formally extended to strong disorder, they suggested that there can 
exist the so-called {\it Bose glass} phase, in the sense that there is a Bose-Einstein
condensate, $n_0\neq 0$, but there is no superfluidity, $n_s\equiv 0$. 
Weakly interacting Bose gas in the presence of disorder was also theoretically
studied in Refs.~\cite{Lopatin,Paul,Modugno,Bilas,Sanchez-Palencia,Lugan,Falco}
and the experiments demonstrating the localization
of atomic matter waves were performed recently~\cite{Lye,Clement,Fort}.

The arising Bose glass phase, if any, would be of high importance for 
the experiments with $^4$He-filled porous media~\cite{39}.
A porous material can be mimicked well by an external random potential. 
This is because each pore represents an external local potential. At the 
same time, since pores enjoy random properties, being of different sizes, 
shapes, and being randomly distributed in space, they do form for a Bose 
system a kind of a spatially random potential~\cite{39}. However, Monte 
Carlo simulations~\cite{40} as well as numerical calculations in the frame 
of the random-phase approximation~\cite{41}, accomplished for a Bose system 
with strong disorder, though with asymptotically weak interactions, 
revealed no Bose glass phase. But maybe this phase could appear 
when both disorder and interactions were strong?
In the present paper we shall give our answer to this question.

In Ref.~\cite{38}, a self-consistent stochastic mean-field approximation
for Bose systems in external random potentials,
allowing the treatment of arbitrarily strong interactions and 
arbitrarily strong disorder, have been developed.
It has been found that the disordered system 
contains several particle fractions whose relative size between $0$ and $1$
defines the main system properties.
There exists the fraction $n_0$ of condensed atoms 
and there is the fraction $n_N$ of normal uncondensed atoms. There 
also appears the fraction $n_G$ of a glassy component. An important role 
is played by the anomalous average $\sigma$, whose absolute value $|\sigma|$ 
quantifies the relative density of pair-correlated atoms~\cite{42}. Finally, 
there exists the superfluid fraction $n_s$. All these quantities are 
defined by the solution of a system of nonlinear equations,
whose exact analysis would require numerical calculations.
It is the aim of the present paper to give an analysis of the behavior
of these various fractions under varying strengths of interactions and disorder
in the domain where $n_0\ne0$.
Throughout the paper we use the system of units, where $\hbar\equiv 1$ and $k_B\equiv 1$.

\section{Main Definitions and Notations}

First, we must define the system to be considered. The Hamiltonian 
energy operator has the standard form
\begin{eqnarray}
\hat H 
&=&
\int \psi^\dagger({\bf r}) 
\left [ - \frac{\nabla^2}{2m} +
\xi({\bf r}) \right ]
\psi({\bf r})
\; d{\bf r}
\nonumber\\
&+&
\frac{\Phi_0}{2}
\int \psi^\dagger({\bf r}) \psi^\dagger({\bf r}) \psi({\bf r})\psi({\bf r})\; d{\bf r} 
\;,
\label{1}
\end{eqnarray}
in which $\psi^\dagger({\bf r})$ is the Bose field operator, $\xi({\bf r})$ is 
an external random potential, and the interaction strength is
\begin{equation}
\label{2}
\Phi_0 = 4\pi
\,
\frac{a_s}{m}
\;,
\end{equation}
with the scattering length $a_s$ and the atomic mass $m$.
The random  potential is centered around zero, so that its stochastic average vanishes,
\begin{equation}
\label{3}
\langle\langle
\xi({\bf r})
\rangle\rangle = 0
\;.
\end{equation}
The stochastic correlation function has the general form
\begin{equation}
\label{4}
\langle\langle
\xi({\bf r}) \xi({\bf r}') 
\rangle\rangle
=
R({\bf r}-{\bf r}')
\;.
\end{equation}

For what follows, it is important to distinguish between the stochastic 
averaging over the random-field distribution, which is denoted through 
the angular double brackets
$\langle\langle\ldots\rangle\rangle$,
as in Eqs.~(\ref{3}), (\ref{4}), 
and the statistical averaging over the quantum degrees of freedom, which, 
for an operator $\hat A$, is denoted as
\begin{equation}
\label{8}
\langle \hat A\rangle_H
\equiv
{\rm Tr} \hat\rho \hat A
\; ,
\end{equation}
where $\hat\rho=\hat\rho[H]$ is a statistical operator having the Gibbs 
form with a grand Hamiltonian $H$. The total averaging, given by the 
simple angular brackets $\langle\ldots\rangle$, includes both the stochastic as well 
as the quantum averaging,
\begin{equation}
\label{9}
\langle \hat A\rangle
\equiv 
\langle\langle
\left ( \langle \hat A\rangle_H \right )
\rangle\rangle
=
\langle\langle
{\rm Tr} \hat\rho \hat A
\rangle\rangle
\;.
\end{equation}

The appearance of a Bose-Einstein condensate implies the spontaneous gauge 
symmetry breaking, which can be realized by the Bogolubov shift~\cite{43,44} 
of the field operator
\begin{equation}
\label{10}
\psi({\bf r}) \rightarrow \hat\psi({\bf r}) \equiv \eta({\bf r}) + \psi_1({\bf r}) \; ,
\end{equation}
where $\eta({\bf r})=\langle \hat\psi({\bf r}) \rangle$
is the condensate wave function and $\psi_1({\bf r})$ 
is the Bose field operator of uncondensed atoms. The field variables 
$\eta({\bf r})$ and $\psi_1({\bf r})$ are treated as two independent variables,
orthogonal to each other,
\begin{equation}
\label{11}
\int \eta^*({\bf r}) \psi_1({\bf r})\; d{\bf r} = 0 \; ,
\end{equation}
which excludes the double counting of the degrees of freedom. The 
quantum-number conservation condition
\begin{equation}
\label{12}
\langle \psi_1({\bf r}) \rangle = 0 
\end{equation}
defines $\eta({\bf r})$ as the system order parameter, equal to 
the average $\langle\hat\psi({\bf r})\rangle$.

These two field variables obey two normalization conditions. The 
condensate function is normalized to the number of condensed atoms
\begin{equation}
\label{13}
N_0 = \int |\eta({\bf r})|^2 \; d{\bf r} \; ,
\end{equation}
while the number of uncondensed atoms
\begin{equation}
\label{14}
N_1 = \langle \hat N_1 \rangle
\end{equation}
normalizes the number operator
\begin{equation}
\label{15}
\hat N_1 \equiv \int \psi_1^\dagger ({\bf r}) \psi_1({\bf r}) \; d{\bf r}
\end{equation}
for uncondensed atoms. The system stability is guaranteed by minimizing 
the grand thermodynamic potential
\begin{equation}
\label{16}
\Omega = - T
\langle\langle
\ln {\rm Tr} e^{-\beta H}
\rangle\rangle
\end{equation}
under the constraints of the two normalization conditions~(\ref{13}) and~(\ref{14}), 
which requires the use of two Lagrange multipliers, so that the grand 
effective Hamiltonian in Eq.~(\ref{16}) is
\begin{equation}
\label{17}
H = \hat H - \mu_0 N_0 - \mu_1 \hat N_1
\;,
\end{equation}
with $\hat H$ being the energy operator~(\ref{1}) under the Bogolubov shift~(\ref{10}).
The form of the grand potential~(\ref{16}) corresponds to the quenched 
disorder.

In view of the zero-centered random potential, satisfying Eq.~(\ref{3}), the 
system can be considered as uniform on average, such that
\begin{equation}
\label{18}
\eta({\bf r})
=
\sqrt{\rho_0}
\qquad
\left(
    \rho_0
    \equiv
    N_0/V 
\right)
\;.
\end{equation}
The field operator of uncondensed atoms can be expanded over plane 
waves,
\begin{equation}
\label{19}
\psi_1({\bf r}) = \frac{1}{\sqrt{V}} \sum_{k\neq 0}
a_k e^{i{\bf k}\cdot{\bf r}} \; .
\end{equation}
The operators $a_k$ in the momentum representation define the normal 
average
\begin{equation}
\label{20}
n_k
\equiv
\langle a_k^\dagger a_k \rangle
\end{equation}
and the anomalous average
\begin{equation}
\label{21}
\sigma_k
\equiv
\langle a_k a_{-k} \rangle 
\;.
\end{equation}

The major quantities to be studied are the densities of different 
components. The condensate density $\rho_0=\rho-\rho_1$ is expressed 
through the total average density
\begin{equation}
\label{22}
\rho
\equiv
N/V
=
\rho_0 + \rho_1
\;,
\end{equation}
where $N$ is the total number of atoms, and through the density of 
uncondensed atoms
\begin{equation}
\label{23}
\rho_1
\equiv
\frac{1}{V}
\sum_{k\neq 0}
n_k
\;.
\end{equation}
The anomalous average
\begin{equation}
\label{24}
\sigma_1
\equiv
\frac{1}{V}
\sum_{k\neq 0}
\sigma_k
\end{equation}
gives the density $|\sigma_1|$ of pair-correlated atoms~\cite{42}. In the presence 
of random fields, there appears an additional important quantity, the density 
of the glassy component
\begin{equation}
\label{25}
\rho_G
\equiv
\frac{1}{V}
\sum_{k\neq 0}
\langle\langle
\left|
    \langle a_k \rangle_H
\right|^2
\rangle\rangle
\;,
\end{equation}
whose definition is analogous to the Edwards-Andersen order parameter 
for spin glasses~\cite{45}.

Superfluidity is characterized by the superfluid fraction
\begin{equation}
\label{26}
n_s \equiv \frac{1}{3mN}\; \lim_{v\rightarrow 0} \; \frac{\partial}{\partial{\bf v}}
\;\cdot
\langle \hat{\bf P}_v\rangle_v \; ,
\end{equation}
defined as the fraction of atoms nontrivially responding to the 
velocity boost with the velocity ${\bf v}$, under $v\equiv|{\bf v}|\rightarrow 0$. 
Here
\begin{equation}
\label{27}
\hat{\bf P}_v \equiv \hat{\bf P} + m{\bf v} N \qquad 
\left ( \hat{\bf P} = \sum_k {\bf k} a_k^\dagger a_k \right )
\end{equation}
is the total momentum of the moving system, and the averaging in Eq.~(\ref{26})
implies that with the Hamiltonian of the moving system
\begin{equation}
\label{28}
H_v
\equiv
H
+
\frac{mv^2}{2}
\;
\hat N
+ \sum_k ( {\bf k} \cdot {\bf v})
a_k^\dagger a_k
\;,
\end{equation}
where
\begin{displaymath}
\hat N
=
N_0
+
\sum_{k \ne 0}
a_k^\dagger a_k
\;.
\end{displaymath}
It can be shown (see, e.g., Refs.~\cite{3,38}) that definition~(\ref{26}) yields
\begin{equation}
\label{29}
n_s = 1 - \frac{2Q}{3T}
\;,
\end{equation}
where
\begin{equation}
\label{30}
Q \equiv \frac{\langle\hat{\bf P}^2\rangle}{2mN}
\end{equation}
is the dissipated heat per atom.

\section{Stochastic Mean-Field Approximation}

Since our aim is to treat arbitrarily strong interactions and 
disorder, we cannot neglect any part of the total grand Hamiltonian~(\ref{17}). For 
example, if we would omit the terms of the third and fourth order, containing 
the products of three and four operators $a_k$ or $a_k^\dagger$, as well as the 
third-order term, including the product $a_k^\dagger a_p\xi_{k-p}$, we would come 
to the Bogolubov approximation, used by Huang and Meng~\cite{37}, and many others, 
which allows the consideration of only asymptotically weak interactions and 
disorder. Contrary to this, we shall retain all terms of the Hamiltonian, 
using the stochastic mean-field approximation of Ref.~\cite{38}. This approximation 
was previously shown to give a very accurate description of different 
statistical systems with stochastic fields, as is summarized in Refs.~\cite{46,47,48}.
The stochastic mean-field approximation for Bose systems 
with random fields has been described in full detail in the recent paper~\cite{38}. 
Therefore in the present work, we limit ourselves by mentioning only the 
principal points of this approximation and by reviewing the resulting 
formulas that are necessary for the following analysis.
In the following we shall choose the condensate wave function $\eta$ as real
without restriction of generality.

The third- and fourth-order terms of the Hamiltonian with respect to the 
products of the operators $a_k$ and $a_k^\dagger$ are treated by means of the 
Hartree-Fock-Bogolubov approximation, similarly to the case without random 
fields~\cite{30,31,32,33}. A special care is taken with regard to the third-order term 
containing the random field $\xi_k$. To this end, we use the fact that there 
are two types of averaging, the stochastic averaging, as in Eqs.~(\ref{3}), (\ref{4}), 
and~(\ref{7}), and the quantum statistical averaging, as defined in Eq.~(\ref{8}). Let 
us introduce the random quantity
\begin{equation}
\label{31}
\alpha_k \equiv \langle a_k\rangle_H \; ,
\end{equation}
which is the quantum average~(\ref{8}) of the operator $a_k$. 
The random variable~(\ref{31}) is not $0$, though its stochastic average
\begin{displaymath}
\langle\langle
\alpha_k
\rangle\rangle
=
\langle a_k \rangle
=
0
\end{displaymath}
is zero because of condition~(\ref{12}). Then, using notation~(\ref{31}), we accomplish 
a mean-field-type decoupling for the third-order term
\begin{equation}
\label{32}
a_k^\dagger a_p \xi_{k-p} = \left ( \alpha_k^* a_p + a_k^\dagger \alpha_p -
\alpha_k^* \alpha_p \right ) \xi_{k-p} \; ,
\end{equation}
where only the quantum averaging is involved, but no stochastic averaging is 
taken. Keeping here the {\it stochastic} averaging unapproximated makes it possible to 
consider any strength of disorder.

The following important step is the use of the nonuniform and {\it nonlinear}, 
with respect to the random variable $\xi_k$, canonical transformation
\begin{equation}
\label{33}
a_k = u_k \hat b_k + v_{-k}^* \hat b_{-k}^\dagger - 
\frac{\varphi_k}{\omega_k+mc^2} \; ,
\end{equation}
containing a new random variable $\varphi_k$ to be defined later. Here,
\begin{eqnarray}
u_k^2 &=& \frac{\omega_k+\varepsilon_k}{2\varepsilon_k} \; , \qquad v_k^2 =
\frac{\omega_k-\varepsilon_k}{2\varepsilon_k} \; , \qquad
\nonumber\\
\omega_k &=& \frac{k^2}{2m} + mc^2 \; , \qquad 
\varepsilon_k^2 = \omega_k^2 - (mc^2)^2 \; .
\label{34}
\end{eqnarray}
The latter equation, for $\varepsilon_k$, can be represented as the Bogolubov 
spectrum
\begin{equation}
\label{35}
\varepsilon_k =\sqrt{ (ck)^2 + \left ( \frac{k^2}{2m}\right )^2} \; ,
\end{equation}
however with the sound velocity $c$ differing from that of the Bogolubov 
form. Rather, $c$ is given here as the solution to the equation
\begin{equation}
\label{36}
mc^2 = (\rho_0+\sigma_1) \Phi_0 \; .
\end{equation}

The random variable $\varphi_k$, introduced in the canonical transformation~(\ref{33}), 
has to be chosen so that to simplify the total Hamiltonian. If the variable 
$\varphi_k$ satisfies the Fredholm equation
\begin{equation}
\label{37}
\varphi_k = \frac{\sqrt{N_0}}{V} \; \xi_k - 
\frac{1}{V} \sum_{p\neq 0} 
\frac{\xi_{k-p}\varphi_p}{\omega_p+mc^2} \; ,
\end{equation}
then the Hamiltonian acquires the simple form
\begin{equation}
\label{38}
H = E_B + \sum_{k\neq 0} \varepsilon_k \hat b_k^\dagger \hat b_k +
\sqrt{N_0}\; \varphi_0 \; ,
\end{equation}
in which the quantum variables $\hat b_k$ and $\hat b_k^\dagger$ are separated 
from the random variable $\varphi_k$, and where the first term is a $c$-number
quantity
\begin{displaymath}
E_B = \frac{1}{2} \sum_{k\neq 0} (\varepsilon_k - \omega_k)  -
\frac{N}{2} \left [ 2 \left ( \rho^2 - \rho_0^2\right ) +
(\rho_0 +\sigma_1)^2 \right ] \Phi_0 \; .
\end{displaymath}
The last term in Hamiltonian~(\ref{38}) is obtained by using Eq.~(\ref{37}),
with the expression for $\varphi_0$ defined by the equation
\begin{displaymath}
\varphi_0
=
\frac{\sqrt{N_0}}{V}
\xi_0
-
\frac{1}{V}
\sum_{p\ne 0}
\frac{\xi_p^* \varphi_p}{\omega_p+mc^2}
\;.
\end{displaymath}
The relation between the random variable $\alpha_k$, defined in Eq.~(\ref{31}), and 
the random variable $\varphi_k$, satisfying Eq.~(\ref{37}), follows from Eq.~(\ref{33}), from 
where
\begin{equation}
\label{39}
\alpha_k = - \frac{\varphi_k}{\omega_k+mc^2} \; .
\end{equation}

With Hamiltonian~(\ref{38}), it is straightforward to calculate the normal average~(\ref{20}),
which gives
\begin{equation}
\label{40}
n_k = \frac{\omega_k}{2\varepsilon_k}
\;
{\rm coth}
\left ( 
\frac{\varepsilon_k}{2T} \right ) - \frac{1}{2} + 
\langle\langle
|\alpha_k|^2
\rangle\rangle
\;,
\end{equation}
and the anomalous average~(\ref{21}), yielding
\begin{equation}
\label{41}
\sigma_k
=
-\frac{mc^2}{2\varepsilon_k} \;
{\rm coth}
\left( 
    \frac{\varepsilon_k}{2T}
\right)
+
\langle\langle
|\alpha_k|^2
\rangle\rangle
\;.
\end{equation}
The last terms in Eqs.~(\ref{40}) and (\ref{41}) are caused by the random potential. 
According to relation (\ref{39}), we have
\begin{equation}
\label{42}
\langle\langle
|\alpha_k|^2
\rangle\rangle
=
\frac{\langle\langle|\varphi_k|^2\rangle\rangle}{(\omega_k+mc^2)^2}
\;.
\end{equation}
The density of uncondensed atoms~(\ref{23}) consists of two terms,
\begin{equation}
\label{43}
\rho_1  = \rho_N + \rho_G \; ,
\end{equation}
the first of which is the density of normal uncondensed atoms
\begin{equation}
\label{44}
\rho_N = \frac{1}{2}
\int
\left [ \frac{\omega_k}{\varepsilon_k}\;
{\rm coth}\left ( \frac{\varepsilon_k}{2T}\right ) - 1 \right ]
\frac{d{\bf k}}{(2\pi)^3}
\end{equation}
and the second term is the density~(\ref{25}) of the glassy component
\begin{equation}
\label{45}
\rho_G = \int \frac{\langle\langle|\varphi_k|^2\rangle\rangle}{(\omega_k+mc^2)^2}
\;
\frac{d{\bf k}}{(2\pi)^3}
\;.
\end{equation}
The anomalous average~(\ref{24}) also is a sum
\begin{equation}
\label{46}
\sigma_1 = \sigma_N + \rho_G 
\end{equation}
of the term
\begin{equation}
\label{47}
\sigma_N = - \frac{1}{2}\int \frac{mc^2}{\varepsilon_k}\;
{\rm coth}\left ( \frac{\varepsilon_k}{2T}\right ) 
\frac{d{\bf k}}{(2\pi)^3}
\end{equation}
and of the glassy density~(\ref{45}).

The superfluid fraction~(\ref{29}) is expressed through the dissipated 
heat~(\ref{30}). For the latter, we find
\begin{equation}
\label{48}
Q = Q_N + Q_G \; ,
\end{equation}
where the first term
\begin{equation}
\label{49}
Q_N = \frac{1}{8m\rho} \int 
\frac{k^2}{{\rm sinh}^2(\varepsilon_k/2T)} \; \frac{d{\bf k}}{(2\pi)^3} 
\end{equation}
is the heat dissipated by normal uncondensed atoms and the second 
term
\begin{equation}
\label{50}
Q_G = \frac{1}{2m\rho} \int 
\frac{k^2\langle\langle|\varphi_k|^2\rangle\rangle}{\varepsilon_k(\omega_k+mc^2)}
\;
{\rm coth}
\left ( \frac{\varepsilon_k}{2T} \right )
 \frac{d{\bf k}}{(2\pi)^3} 
\end{equation}
is the heat dissipated by the glassy component caused by the 
random potential.

\section{$\delta$-Correlated Random Potential}

To proceed further in practical calculations, we must specify the 
type of random potential. For this purpose, we take the Gaussian $\delta$-correlated
random potential with the local correlation function
\begin{equation}
\label{5}
R({\bf r}) = R_0\; \delta({\bf r}) \; .
\end{equation}
Then, by means of the Fourier transformation
\begin{equation}
\label{6}
\xi({\bf r}) = \frac{1}{V} \sum_k \xi_k e^{i{\bf k}\cdot{\bf r}} \; ,
\end{equation}
the stochastic correlator~(\ref{4}) reduces to
\begin{equation}
\label{7}
\langle\langle 
\xi_k^* \xi_p
\rangle\rangle
=
\delta_{kp} R_0 V 
\;,
\end{equation}
where $V$ is the system volume.
The calculation of the stochastic average $\langle\langle|\varphi_k|^2\rangle\rangle$,
involving the method of self-similar factor approximants~\cite{49,50},
can be done as has been thoroughly explained in Ref.~\cite{38}.

For what follows, it is convenient to deal with dimensionless quantities. 
We introduce the notation for the condensate fraction
\begin{equation}
\label{51}
n_0
\equiv
\rho_0/\rho
\;,
\end{equation}
the normal fraction of uncondensed atoms
\begin{equation}
\label{52}
n_N
\equiv
\rho_N/\rho
\; ,
\end{equation}
the glassy fraction
\begin{equation}
\label{53}
n_G
\equiv
\rho_G/\rho
\; ,
\end{equation}
and for the dimensionless anomalous average
\begin{equation}
\label{54}
\sigma
\equiv
\sigma_N/\rho
\;.
\end{equation}

The interaction strength is characterized by the gas parameter
\begin{equation}
\label{55}
\gamma \equiv \rho^{1/3} a_s \; .
\end{equation}
The dimensionless temperature is
\begin{equation}
\label{56}
t
\equiv
mT/\rho^{2/3}
\;.
\end{equation}
The strength of disorder is quantified by the disorder parameter
\begin{equation}
\label{57}
\nu
\equiv
\frac{7m^2R_0}{4\pi\rho^{1/3}}
\; .
\end{equation}
Finally, we define the dimensionless sound velocity
\begin{equation}
\label{58}
s
\equiv
mc/\rho^{1/3}
\; .
\end{equation}

We shall consider the case $n_0\ne 0$, where the gauge symmetry is spontaneously broken.
The case of the unbroken gauge symmetry with $n_0=0$ will be considered in future work.
The sound velosity $s$ is then well-defined and given by the solution to Eq.~(\ref{36}),
which in the dimensionless
quantities, and remembering expression~(\ref{2}) for the interaction strength, 
takes the form
\begin{equation}
\label{59}
s^2 = 4\pi\gamma ( 1 - n_N + \sigma) \; .
\end{equation}

The condensate fraction~(\ref{51}) can be found from Eqs.~(\ref{22}) and~(\ref{43}), which 
can be reduced to the equation
\begin{equation}
\label{60}
n_0 + n_N + n_G = 1 \; .
\end{equation}
For the normal fraction~(\ref{52}), taking into account Eq.~(\ref{44}), we have,
for $n_0\ne 0$,
\begin{eqnarray}
n_N
&=&
\frac{s^3}{3\pi^2}
\left\{
    1 
    +
    \frac{3}{2\sqrt{2}} 
    \int_0^\infty
    \left(
        \sqrt{1+x^2}-1
    \right )^{1/2}
\right.
\nonumber\\
&&
\times
\left.
    \left[
        {\rm coth}\left ( \frac{s^2x}{2t}\right ) - 1
    \right]
\; dx 
\right\}
\;.
\label{61}
\end{eqnarray}
The anomalous average~(\ref{54}), with Eq.~(\ref{47}) and the dimensional 
regularization~\cite{5,38}, becomes
\begin{eqnarray}
\sigma 
&=&
\frac{2s^2}{\pi^{3/2}} \; \sqrt{\gamma n_0} -
\frac{s^3}{2\sqrt{2}\pi^2} \int_0^\infty
\frac{(\sqrt{1+x^2}-1)^{1/2}}{\sqrt{1+x^2}}
\nonumber\\
&&
\times
\left [
{\rm coth}\left ( \frac{s^2x}{2t} \right ) - 1 
\right ]\; dx  \; .
\label{62}
\end{eqnarray}
For the glassy fraction~(\ref{53}), we obtain from Eq.~(\ref{45})
\begin{equation}
\label{63}
n_G = \frac{\nu(1-n_N)}{\nu+7s^{4/7}(s-\nu)^{3/7}} \; .
\end{equation}
Finally, for the superfluid fraction~(\ref{29}), employing Eqs.~(\ref{30}), 
(\ref{48}), (\ref{49}), and (\ref{50}), we find
\begin{eqnarray}
\label{64}
n_s &=& 1 - \frac{4}{3}\; n_G
\\
&-& 
\frac{s^5}{6\sqrt{2}\pi^2 t}
\int_0^\infty
\frac{x(\sqrt{1+x^2}-1)^{3/2}\;dx}{\sqrt{1+x^2}\;{\rm sinh}^2(s^2x/2t)}\; .
\nonumber
\end{eqnarray}

In the following we shall concentrate on the ground-state properties of the system, 
corresponding to the limit of zero temperature $t\to 0$. In this case, the Bose 
system without disorder would be completely superfluid, $n_s=1$, but the condensate 
fraction is depleted by interactions. External random fields deplete both 
the condensate and superfluid fractions, so that $n_0<1$ and $n_s<1$.
What would be the behavior of these 
fractions when varying the interaction and disorder strengths, that is, the 
gas parameter (\ref{55}) and the disorder parameter (\ref{57})?

At zero temperature, the normal fraction~(\ref{61}) reduces to
\begin{equation}
\label{65}
n_N = \frac{s^3}{3\pi^2} \; ,
\end{equation}
while the anomalous fraction~(\ref{62}) becomes
\begin{equation}
\label{66}
\sigma= \frac{2s^2}{\pi^{3/2}} \; \sqrt{\gamma n_0} \; ,
\end{equation}
with $n_G$ and $n_0$ being defined from Eq.~(\ref{63}) and normalization~(\ref{60}),
respectively.
The superfluid fraction~(\ref{64}) at zero temperature takes the form
\begin{equation}
\label{67}
n_s =  1 - \frac{4}{3}\; n_G \; .
\end{equation}
Equations~(\ref{59}), (\ref{60}), (\ref{65}), (\ref{66}), and (\ref{67})
determine their solutions as functions of two variables, the gas parameter~(\ref{55})
and the disorder parameter~(\ref{57}).

Let us, first, consider the asymptotic behavior of the solutions. If 
the disorder parameter $\nu$ is finite and $\gamma$ tends to zero, there are 
no physical solutions, which corresponds to the stochastic instability 
of the ideal Bose-condensed gas~\cite{38}.

For a finite disorder parameter and very strong interactions, such 
that $\gamma\rightarrow\infty$, we have
\begin{equation}
\label{68}
s \simeq s_\infty - \frac{1}{64} \left (
\frac{\pi^5}{9}\right )^{1/3} \left [ 1 +
\frac{\nu}{7s^{4/7}_\infty(s_\infty - \nu)^{3/7}} \right ]
\frac{1}{\gamma^3} \; ,
\end{equation}
with the limit
\begin{equation}
\label{69}
s_\infty \equiv \left ( 3\pi^2 \right )^{1/3} \; .
\end{equation}
Then the normal fraction~(\ref{65}) becomes
\begin{equation}
\label{70}
n_N \simeq 1- \frac{\pi}{64} \left [ 1 +
\frac{\nu}{7s_\infty^{4/7}(s_\infty-\nu)^{3/7}}\right ]
\frac{1}{\gamma^3} \; ,
\end{equation}
with the anomalous fraction~(\ref{66}) being
\begin{equation}
\label{71}
\sigma \simeq \frac{(9\pi)^{1/3}}{4\gamma} + O\left ( \frac{1}{\gamma^4}
\right ) \; .
\end{equation}
For the glassy fraction~(\ref{63}), we find
\begin{equation}
\label{72}
n_G \simeq \frac{\pi\nu}{448s_\infty^{4/7}(s_\infty-\nu)^{3/7}}
\left ( \frac{1}{\gamma^3}\right ) \; .
\end{equation}
Normalization~(\ref{60}) gives the condensate fraction
\begin{equation}
\label{73}
n_0 \simeq \frac{\pi}{64\gamma^3} + O\left ( 
\frac{1}{\gamma^4}\right ) \; .
\end{equation}
Equation~(\ref{67}) yields the superfluid fraction
\begin{equation}
\label{74}
n_s \simeq 1 - 
\frac{\pi\nu}{336s_\infty^{4/7}(s_\infty-\nu)^{3/7}}
\left ( \frac{1}{\gamma^3}\right ) \; .
\end{equation}
As we see, strong interactions tend to destroy a Bose-Einstein condensate, 
suppressing $n_0$, but increase the superfluid fraction $n_s$. In this 
limit, $n_s\gg n_0$.

\begin{figure}[t]
\centering

  \includegraphics[width=8cm]{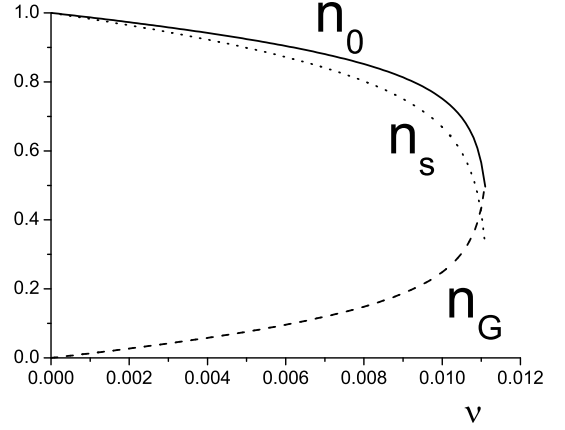}

\caption{Condensate fraction $n_0$ (solid line), superfluid fraction
$n_s$ (dotted line), and glassy fraction $n_G$ (dashed line) as functions of
the dimensionless disorder parameter $\nu$ for very weak interactions, with the
gas parameter $\gamma=10^{-5}$.
        }
\label{Fig1}
\end{figure}

\begin{figure}[t]
\centering

  \includegraphics[width=8cm]{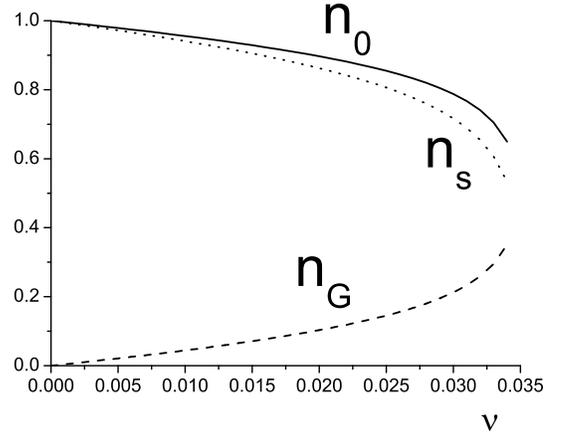}

\caption{The same fractions, $n_0$, $n_s$, and $n_G$, as in Figure~\ref{Fig1}, but
for slightly larger interactions, with the gas parameter $\gamma=0.001$. Here 
the superfluid fraction becomes smaller than the condensate fraction, when 
disorder increases, but $n_s$ is never smaller than the glassy fraction 
$n_G$.
        }
\label{Fig2}
\end{figure}

\begin{figure}[t]
\centering

  \includegraphics[width=8cm]{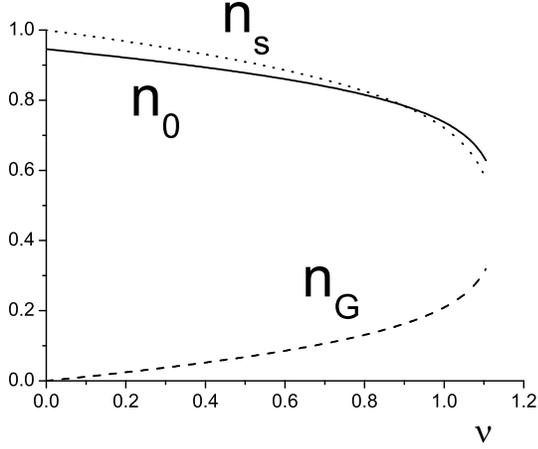}

\caption{Fractions $n_0$, $n_s$, and $n_G$ for the gas parameter
$\gamma=0.1$. The superfluid fraction becomes lower than the condensate
fraction at the disorder parameter $\nu\approx 0.9$.
        }
\label{Fig3}
\end{figure}

\begin{figure}[t]
\centering

  \includegraphics[width=8cm]{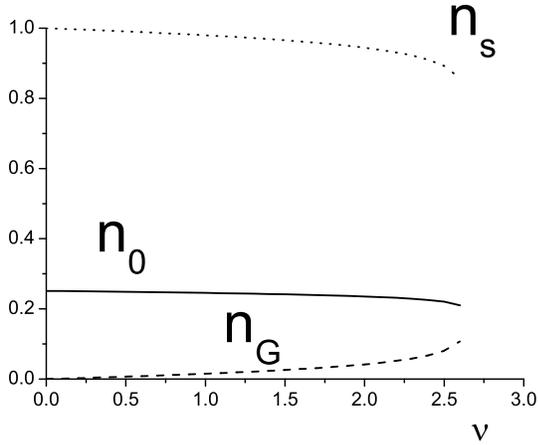}

\caption{Fractions $n_0$, $n_s$, and $n_G$ for the gas parameter $\gamma=0.5$. 
The superfluid fraction is always higher than the condensate fraction.
        }
\label{Fig4}
\end{figure}

\begin{figure}[t]
\centering

  \includegraphics[width=8cm]{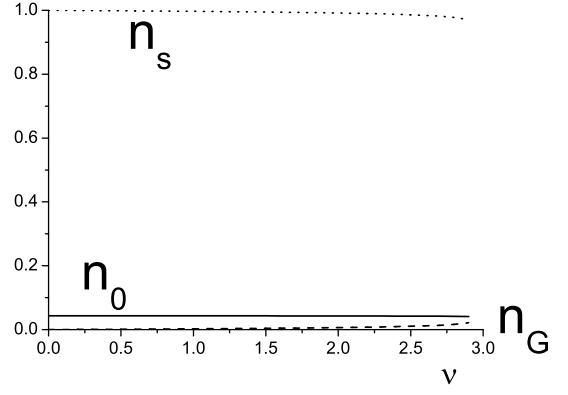}

\caption{Fractions $n_0$, $n_s$, and $n_G$ at a relatively large gas
parameter $\gamma=1$. The condensate and glassy fractions are strongly
suppressed.
        }
\label{Fig5}
\end{figure}

\begin{figure}[t]
\centering

  \includegraphics[width=8cm]{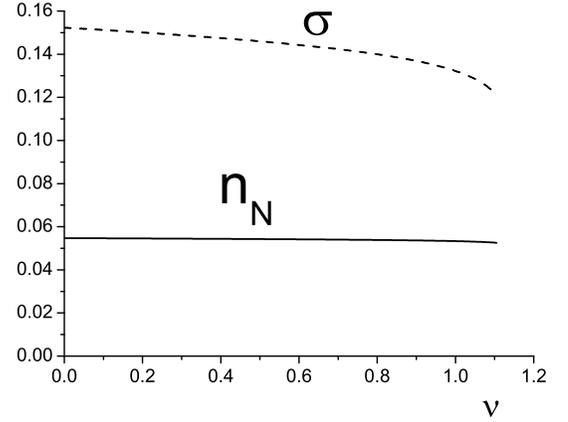}

\caption{Normal fraction $n_N$ (solid line) and anomalous fraction $\sigma$
(dashed line) as functions of the disorder parameter $\nu$ for weak interactions,
with the gas parameter $\gamma=0.1$. The anomalous fraction $\sigma$ is about
three times larger than the normal fraction $n_N$.
        }
\label{Fig6}
\end{figure}

\begin{figure}[t]
\centering

  \includegraphics[width=8cm]{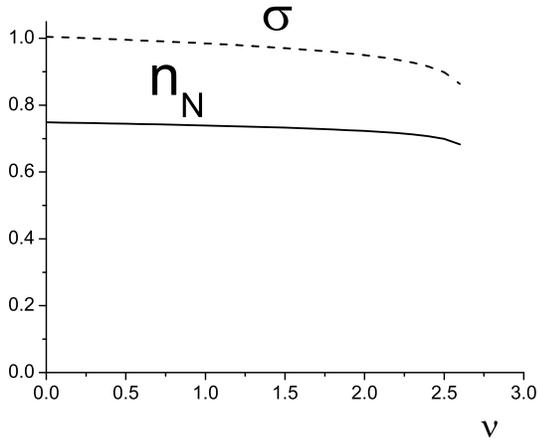}

\caption{Normal, $n_N$, and anomalous, $\sigma$, fractions as functions of
the disorder parameter $\nu$ for the intermediate strength, with $\gamma=0.5$. The
values of $\sigma$ and $n_N$ are close to each other, though, $\sigma>n_N$.
        }
\label{Fig7}
\end{figure}

\begin{figure}[t]
\centering

  \includegraphics[width=8cm]{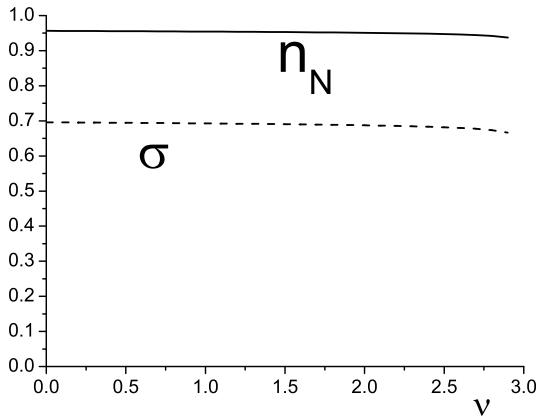}

\caption{Normal, $n_N$,  and anomalous, $\sigma$, fractions for rather strong 
interactions, with the gas parameter $\gamma=1$. Here, contrary to 
Figs.~\ref{Fig6}
and \ref{Fig7}, 
the normal fraction becomes larger than the anomalous one, though they are 
close to each other.
        }
\label{Fig8}
\end{figure}

\begin{figure}[t]
\centering

  \includegraphics[width=8cm]{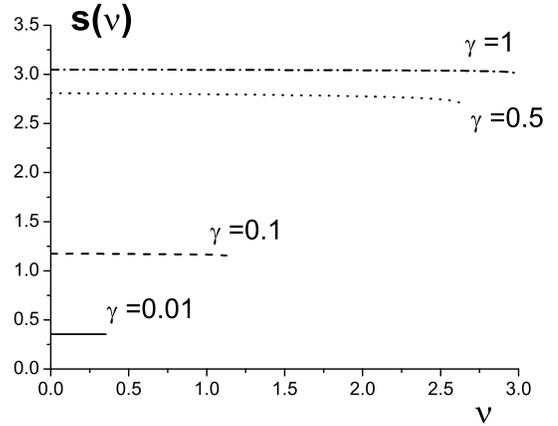}

\caption{Dimensionless sound velocity $s$ as a function of the disorder 
parameter $\nu$ for different gas parameters. Each line is marked by the 
corresponding $\gamma$.
        }
\label{Fig9}
\end{figure}

\begin{figure}[t]
%
%
%
%

  \hspace{-3cm}\includegraphics[width=9cm]{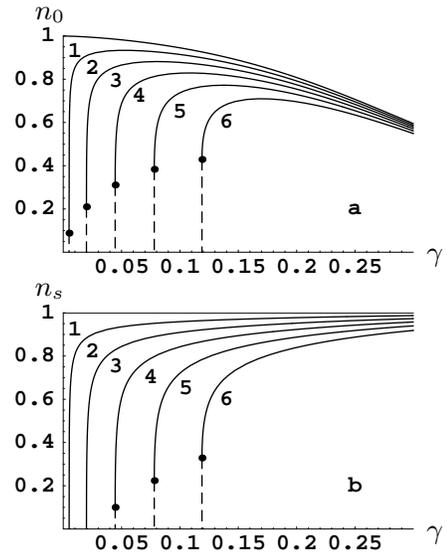}

\caption{Condensate fraction (a) and superfluid fraction (b)
         as functions of the gas parameter $\gamma$
         for $\nu=$0~(1), 0.25~(2), 0.5~(3), 0.75~(4), 1~(5), 1.25~(6).
        }
\label{csg}
\end{figure}

\begin{figure}[t]
%
%

  \hspace{-3cm}\includegraphics[width=9cm]{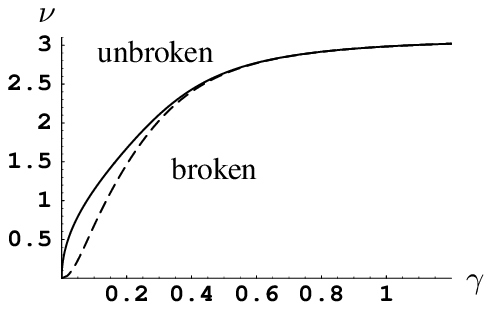}

\caption{Phase diagram in the $(\gamma,\nu)$-plane.
The solid line is the line of the first-order quantum phase transition
between the phases of broken and unbroken gauge symmetry.
The dashed line indicates the values of $\nu$ and $\gamma$
at which the condensate fraction takes the maximal value (see Fig.~\ref{csg}a).
        }
\label{Fig10}
\end{figure}

If the gas parameter is kept finite, but the disorder strength is getting 
asymptotically weak, such that $\nu\rightarrow 0$, then the sound velocity is
\begin{equation}
\label{75}
s\simeq s_0(1 - b\nu) \; ,
\end{equation}
where the limit
\begin{equation}
\label{76}
s_0 \equiv (3\pi^2)^{1/3} a
\end{equation}
depends on the interaction strength through the equation
\begin{equation}
\label{77}
a^3 + \frac{(9\pi)^{1/3}}{4\gamma}\; a^2 \left ( 1 - 
\frac{8\gamma^{3/2}}{\sqrt{\pi}}\; \sqrt{1-a^3}\right ) =  1
\end{equation}
and the value of $b$ is given by the equation
\begin{equation}
\label{78}
b\left [ \left ( 2+a^3 \right )\sqrt{1-a^3} +
\frac{3\sqrt{\gamma}}{\pi^{3/2}}\; a^3 s_0^2 \right ] =
\frac{\sqrt{\gamma}}{7\pi^2}\left ( 1 - a^3\right ) s_0 \; .
\end{equation}

The solutions to Eqs.~(\ref{77}) and (\ref{78}) for weak interactions
($\gamma\rightarrow 0$) are
\begin{eqnarray}
\label{79}
a
&\simeq&
\frac{2}{(9\pi)^{1/6}}\; \gamma^{1/2} + 
\frac{16}{(9\pi)^{2/3}}\; \gamma^2
\; ,
\nonumber\\
b
&\simeq&
\frac{1}{7\pi^{3/2}}\; \gamma + O(\gamma^{5/2})
\end{eqnarray}
and for strong interactions ($\gamma\rightarrow\infty$), Eqs.~(\ref{77}) and (\ref{78}) yield
\begin{eqnarray}
\label{80}
a
&\simeq&
1 -  \frac{\pi}{192\gamma^3}
\;,
\nonumber\\
b
&\simeq&
\frac{1}{1344(9\pi)^{1/6}}\left ( \frac{1}{\gamma^3}\right )
+ O \left ( \frac{1}{\gamma^5}\right )
\; .
\end{eqnarray}
With the interaction strength varying from zero to infinity, the value 
of $a$ increases from $0$ to $1$, so that $0\leq a<1$. But the value of 
$b$ remains small for all interactions, $b\ll 1$. For finite $\gamma$, but 
asymptotically weak disorder, when $\nu\rightarrow 0$, the normal fraction is
\begin{equation}
\label{81}
n_N \simeq a^3 (1 - 3b\nu)
\;,
\end{equation}
while the anomalous fraction  becomes
\begin{equation}
\label{82}
\sigma \simeq \frac{2s_0^2}{\pi^{3/2}} \; \sqrt{\gamma(1-a^3)} \left [ 1
- \frac{1-a^3+(28-49 a^3)bs_0}{14s_0}\; \nu\right ] \; ,
\end{equation}
where $a$ and $b$ are the same as above. For the glassy fraction, we have
\begin{equation}
\label{83}
n_G \simeq \frac{1-a^3}{7s_0}\; \nu \; .
\end{equation}
The condensate fraction behaves as
\begin{equation}
\label{84}
n_0 \simeq 1 - a^3 - \frac{1-a^3-21a^3bs_0}{7s_0}\; \nu \; .
\end{equation}
For the superfluid fraction, we obtain
\begin{equation}
\label{85}
n_s\simeq 1 - \frac{4(1-a^3)}{21s_0}\; \nu \; .
\end{equation}
These formulas, to leading order in $\gamma$ and $\nu$,
coinside with the results of Ref.~\cite{37}.
In the absence of disorder, the whole system would be superfluid, but never 
completely condensed for any finite interactions. Consequently, in the limit 
of $\nu\rightarrow 0$, we have $n_s>n_0$.

With increasing disorder, there occurs a first-order phase transition at 
a value $\nu_c=\nu_c(\gamma)$, when the system discontinuously transforms
to a phase with unbroken gauge symmetry. 
At the point $\nu_c$, the fractions $n_0$, $n_s$, $n_G$, and $\sigma$ jump to 
$0$, while $n_N$ jumps to $1$. The behavior of the fractions $n_0$, $n_s$, 
and $n_G$ as functions of the disorder parameter $\nu$ 
for some selected increasing values
of the gas parameter $\gamma$ are shown in 
Figures~\ref{Fig1}-\ref{Fig5}.

The case of very weak interactions, with $\gamma=10^{-5}$, is illustrated in 
Fig.~\ref{Fig1}. According to Eqs.~(\ref{84}) and (\ref{85}), we know that at asymptotically 
weak disorder, when $\nu\rightarrow 0$, the condensate fraction is smaller than 
the superfluid fraction, $n_0<n_s$. But Fig.~\ref{Fig1} shows that with increasing 
disorder the superfluid fraction becomes smaller than the condensate 
fraction. This occurs at a rather small value $\nu$, which is too close to 
$0$ to be noticeable in the figure. Also, we see that disorder suppresses 
the superfluid fraction so that it becomes not merely smaller than $n_0$, 
but eventually even smaller than the glassy fraction $n_G$.

Increasing the interaction as we go through Figs.~\ref{Fig2}-\ref{Fig5}
strengthens superfluidity. Figure~\ref{Fig2}, for $\gamma=0.001$,
demonstrates that, even though there still exists a small value $\nu$, where the 
inequality $n_0<n_s$ changes for $n_s<n_0$, the superfluid fraction remains now 
always larger than the glassy fraction, $n_s>n_G$. 

When increasing interactions further, say, to the gas parameter $\gamma=0.1$, as 
in Fig.~\ref{Fig3}, the point, where the inequality $n_s>n_0$ changes for $n_s<n_0$,
moves to the right, getting closer to the phase transition point $\nu_c$. 
The glassy fraction $n_G$ is always lower than both $n_0$ and $n_s$.

In Fig.~\ref{Fig4}, for $\gamma=0.5$, the superfluid fraction is now always larger than the 
condensate fraction, which distinguishes this figure from the three previous 
ones. The condensate fraction is yet substantially larger than the glassy 
fraction. 

Figure~\ref{Fig5} emphasizes how strong interactions, with $\gamma=1$, favor 
superfluidity, while suppressing both the condensate fraction and the glassy 
fraction. The latter two fractions become rather small, but the superfluid 
fraction is close to one. This also shows that the system can be practically 
completely superfluid, having a tiny condensate fraction, as it happens in
liquid $^4$He. Thus, in Fig.~\ref{Fig5}, the condensate fraction is about $5\%$, 
though the superfluid fraction is almost $100\%$.

The anomalous fraction $\sigma$ and the normal fraction $n_N$ for relatively 
weak interactions, with the gas parameter $\gamma=0.1$, are plotted in Fig.~\ref{Fig6}.
As is seen there, $\sigma$ is about $3$ times larger than $n_N$, which stresses 
the fact that $\sigma$ cannot be neglected.

For the intermediate interaction strength, the normal and anomalous fractions 
are close to each other, as is shown in Fig.~\ref{Fig7} for $\gamma=0.5$. The 
anomalous fraction is yet larger than the normal one.

When interactions become rather strong, as in Fig.~\ref{Fig8} for $\gamma=1$, 
then the normal fraction surpasses the anomalous one. But, anyway, $n_N$ is 
yet close to $\sigma$. Varying the disorder parameter does not have much influence 
on the values of $n_N$ and $\sigma$.

The dimensionless sound velocity $s$ as a function of the disorder parameter 
$\nu$ for different interaction strengths is illustrated in Fig.~\ref{Fig9}. 
As it should be, the larger the gas parameter, the larger is the sound 
velocity. Stronger interactions stabilize the system, increasing the critical 
value $\nu_c$ of the first-order transition. The sound velocity slightly 
diminishes with increasing disorder.

The condensate and superfluid fractions as functions of the interaction 
strength for different disorder parameters are shown in Fig.~\ref{csg}.
As it has been emphasized earlier~\cite{38}, the ideal uniform Bose-condensed 
gas is stochastically unstable. Finite interactions stabilize the system 
against weak disorder. But increasing disorder makes the system unstable, 
when the latter transfers through a first-order phase transition to a 
phase with unbroken gauge symmetry. The jumps of the condensate and 
superfluid fractions in Fig.~\ref{csg} correspond to the phase transition. 
The superfluid fraction increases monotonically with the increasing 
interaction strength. But the remarkable fact is that, for nonvanishing 
disorder parameter, the condensate fraction is not a monotonic function of 
the interaction strength. With increasing gas parameter $\gamma$, the 
condensate fraction first increases, reaches the maximum, and then 
decreases. Thus, there exists the effect of {\it antidepletion}, when the 
increasing interactions result in the {\it rise} of the condensate 
fraction. This effect is due to the presence of disorder, which tends to 
destabilize the system, while the interaction stabilizes it.
The competition between the two tendencies leads to the nonmonotonic behavior
of $n_0$, which is seen in Fig.~\ref{csg}a. The line of the maxima of $n_0$
in the $(\gamma,\nu)$-plane is presented as a dashed line in the phase diagram
in Fig.~\ref{Fig10}.

The line $\nu_c(\gamma)$ of the first-order phase transitions is drawn in
Fig.~\ref{Fig10}. At the point $\gamma=0$, corresponding to the ideal Bose 
gas, the phase transition is of second order. However, the ideal 
Bose-condensed gas is stochastically unstable, and for any infinitesimally 
small disorder parameter $\nu$ it is destroyed, undergoing the phase 
transition to the state with $n_0=0$. Below the line $\nu_c(\gamma)$, there is 
the superfluid phase, with $n_0\neq 0$, $n_s\neq 0$, $n_G\neq 0$, $\sigma
\neq 0$, and $n_N<1$. Above this line, one has the phase of unbroken
gauge symmetry, where 
$n_0=n_s=n_G=\sigma=0$, while $n_N=1$. The limit of $\nu_c(\gamma)$, for 
$\gamma$ tending to infinity, is $(3\pi^2)^{1/3}$. The phase transition 
caused by the increasing disorder is an example of a quantum phase 
transition.

\section{Discussion}

A detailed analysis of the properties of a Bose-condensed system at 
zero temperature in an external random potential has been presented. 
The disorder potential is modelled by the Gaussian uncorrelated disorder. 
The strength of disorder as well as the strength of interactions can be 
arbitrary. The system contains several fractions of particles, the 
condensate fraction $n_0$, superfluid fraction $n_s$, normal fraction 
$n_N$, anomalous fraction $\sigma$, and the fraction of a glassy component 
$n_G$. The behavior of these fractions as functions of the gas parameter 
and the disorder parameter was investigated. The ideal Bose-condensed gas 
is stochastically unstable, since any infinitesimally weak disorder destroys 
it, transferring it to the normal state. Finite interactions stabilize 
the system. Increasing disorder leads to a first-order phase transition 
between the superfluid and normal phases. At asymptotically weak disorder, 
such that $\nu\rightarrow 0$, the superfluid fraction is always larger than the 
condensate fraction, $n_s>n_0$. But increasing disorder, under very weak 
interactions, can invert the latter inequality, when the superfluid 
fraction becomes lower than the condensate fraction, $n_s<n_0$. This is 
in agreement with the Monte Carlos simulations~\cite{40}, where it was noticed 
that sufficiently strong disorder can suppress the superfluid fraction 
making it smaller than the condensate fraction, provided that interaction 
strengths are very weak. However, at sufficiently strong interactions, the 
superfluid fraction gets larger than the condensate fraction for all disorder 
parameters below the phase transition point $\nu_c$. To our knowledge, no
numerical simulations have been accomplished, when both the disorder as well 
as interaction strengths would be strong.

The superfluid and condensate fractions were found always to coexist. It 
may occur that $n_0>n_s$ or $n_0<n_s$, but they are nonzero or zero 
simultaneously. There is no pure Bose glass phase, when $n_s$ would be zero, 
while $n_0$ is nonzero, though the glassy fraction $n_G$, induced by disorder, 
is always present.

Although disorder suppresses superfluidity, $n_s$ never becomes exactly 
zero, as one might conclude from the calculations for asymptotically weak 
disorder~\cite{37}. The pure Bose glass phase does not occur 
in the considered model.

The unusual effect of {\it antidepletion} was found, when increasing 
interactions can increase the condensate fraction in the presence of 
disorder.This effect is caused by the competition
between the disorder destabilizing the system and the interactions, which
stabilize the latter. As a result, in the presence of disorder, the 
condensate fraction becomes a nonmonotonic function of the interaction 
strength.

The change in the behavior of the condensed fraction $n_0$, normal 
fraction $n_N$, glassy fraction $n_G$, and the superfluid fraction $n_s$
results from a competition between the interaction potential and the 
external random potential. These two causes act on the fractions in a 
different way. The increasing interaction always depletes the condensate, 
but increases the superfluid fraction. By depleting the condensate, the 
interaction increases the normal fraction $n_N$. At the same time, 
strengthening disorder increases the glassy fraction and depletes the 
condensate. The competition of all of these, sometimes contradictory,
governs the overall behaviour of the fractions. 

The origin of the phase transition, occurring under the increasing 
disorder, can be understood as follows. Recall that the ideal 
Bose-condensed gas is absolutely unstable with respect to any 
infinitesimally weak random perturbations~\cite{38}. Finite interactions do 
stabilize the Bose-condensed gas. But this stabilization can survive only 
until a finite strength of disorder, when again the system loses stability 
and transforms to the state where the gauge symmetry is not broken.
The point is that disorder 
destroys coherence that is pertinent to Bose-Einstein condensation. By 
destroying coherence, disorder moves the system to a state with no Bose 
condensate.

The calculations in this paper have been done only for zero temperature. 
This is because, first, it has been necessary to understand the behaviour 
of the system under two varying parameters, the interaction strength 
$\gamma$ and the strength of disorder $\nu$. Including temperature makes 
the problem dependent on three parameters. This would essentially complicate 
the consideration making it necessary to resort to mainly numerical 
calculations. We plan to present the details of these calculations in our 
future work. But for low temperatures, the obtained results still do 
hold. Including temperature just leads to more condensate depletion and 
enhancement of the normal fraction.

It is worth emphasizing that when analyzing the behaviour of the 
fractions, we always keep in mind the normalization condition 
(60), according to which the condensate fraction $n_0$, normal fraction 
$n_N$, and the glassy fraction $n_G$ are added to $1$. However, the 
explicit relation between the condensate and superfluid fractions is not 
known, because of which the latter does not enter any simple normalization 
condition, except that $0\leq n_s\leq 1$.

In conclusion, it is important to discuss the possibility of 
experimental observation of the effects described in the present 
paper. Standard experiments are accomplished with trapped atoms. 
The inclusion of a trapping potential in our theory would complicate 
numerical investigation. However, there are two cases, when the 
results of our consideration could be directly applicable to 
experiments. First, the homogeneous picture provides a reasonable 
approximation for wide traps, and, second, it gives a good description 
of the situation at the center of a trap, even if the trap edges are 
rather sharp. This becomes possible because of the known fact that the 
local-density approximation allows for a quite accurate description of 
trapped atoms~\cite{1,2,3,4}, and the uniform case serves as a starting point 
for the local-density approximation.

Keeping in mind the local-density approximation, when close to the 
trap center the system is almost uniform, we must deal with the 
gas of atoms with the positive scattering length, since a homogeneous 
gas with attraction is known to be unstable~\cite{1,2,3,4}. In experiment, 
one can also realize Bose-Einstein condensation of atoms with negative 
scattering length, provided that the atoms are trapped and their number 
does not exceed the critical value $N_c$. A simple formula for the 
critical number $N_c$, giving rather accurate estimates for harmonic 
traps can be represented~\cite{61} as
$$
N_c = \sqrt{\frac{\pi}{2} } \; 
\frac{l_x l_y l_z}{|a_s|\left ( l_x^2 + l_y^2 + l_z^2\right )} \; ,
$$
where $l_\alpha$ is the oscillator length in the $\alpha$-direction 
and the $a_s$ scattering length. A trapped atomic cloud, with a negative 
scattering length, can be stable only when $N<N_c$. This case requires 
a separate investigation. In the present paper, we have considered a 
large system with the number of atoms not bounded from above. This is 
why we have assumed from the beginning that the scattering length is 
positive.

The value of the scattering length can be varied in a wide range, 
for instance, by means of the Feshbach resonance techniques~\cite{4,18}. 
It would be interesting to check in experiment the behavior of the 
system in a fixed random potential, when the interaction strength is 
varied. Such a situation would correspond to Fig.~\ref{csg}. When diminishing 
the scattering length, that is, diminishing the gas parameter $\gamma$, 
we would come to the boundary of stability of the system. Recall that, 
in the absence of interactions, the Bose-condensed system is
stochastically unstable, such that any weak random potential destroys 
the condensate. This phenomenon of stochastic instability was analyzed 
in detail in Ref.~\cite{38}. In order to understand, why this 
phenomenon occurs, it is sufficient to remember that the ideal uniform 
Bose-condensed gas is unstable even in the absence of any random 
potential, which can be easily demonstrated by calculating the system 
compressibility and finding out that the latter diverges in the absence 
of interactions~\cite{4,28}.

Finally, the random potential of the type similar to that considered in the 
present paper can be created in experiment, e.g., by employing the optical 
speckle techniques\cite{Lye,Clement,Fort}. These techniques allow for an 
efficient regulation of the properties of the formed random potential. It is 
possible to organize a frozen random distribution, independent from 
the time variable. It is also feasible to regulate the correlation 
length characterizing the spatial properties of the speckle randomness. 
When the correlation length is much smaller than the healing length,
the effective random potential can be represented as being 
$\delta$-correlated, which has been assumed in the present paper. At the 
same time, we recall that the general theory of Ref.~\cite{38} is applicable 
to random potentials with arbitrary correlation length, although for 
finite-length correlations, calculations would be essentially more 
complicated.
In this way, it looks quite feasible to check the predictions of the 
suggested approach in experiments with atomic Bose gases confined in
wide traps.

\begin{acknowledgments}
This work has been supported by the SFB/TR 12
``Symmetries and universality in mesoscopic systems".
\end{acknowledgments}



\begin{thebibliography}{99}

\bibitem{1}
C.~J.~Pethick and H.~Smith, {\it Bose-Einstein Condensation in Dilute Gases}
(Cambridge University, Cambridge, 2002).

\bibitem{2}
L.~Pitaevskii and S.~Stringari,  {\it Bose-Einstein Condensation} 
(Clarendon, Oxford, 2003).

\bibitem{3}
P.~W.~Courteille, V.~S.~Bagnato, and V.~I.~Yukalov, Laser Phys. {\bf 11},
659 (2001).

\bibitem{4}
V.~I.~Yukalov, Laser~Phys.~Lett. {\bf 1}, 435 (2004).

\bibitem{5}
J.~O.~Andersen, Rev.~Mod.~Phys. {\bf 76}, 599 (2004).

\bibitem{6}
K.~Bongs and K.~Sengstock, Rep.~Prog.~Phys. {\bf 67}, 907 (2004).

\bibitem{7}
V.~I.~Yukalov and M.~D.~Girardeau, Laser~Phys.~Lett. {\bf 2}, 375 (2005).

\bibitem{8}
N.~N.~Bogolubov, J.~Phys.~(Moscow) {\bf 11}, 23 (1947).

\bibitem{9}
N.~N.~Bogolubov, Moscow~Univ.~Phys.~Bull. {\bf 7}, 43 (1947).

\bibitem{10}
M.~H.~Kalos, D.~Levesque, and L.~Verlet, Phys.~Rev.~A {\bf 9}, 2178 (1974).

\bibitem{11}
D.~M.~Ceperley, Rev.~Mod.~Phys. {\bf 67}, 279 (1995).

\bibitem{12}
S.~Giorgini, J.~Boronat, and J.~Casulleras, Phys.~Rev.~A {\bf 60}, 5129 
(1999).

\bibitem{13}
J.~L.~DuBois and H.~R.~Glyde, Phys.~Rev.~A {\bf 63}, 023602 (2001).

\bibitem{14}
C.~C.~Moustakidis and S.~E.~Massen, Phys.~Rev.~A {\bf 65}, 063613 (2002).

\bibitem{15}
J.~L.~DuBois and H.~R.~Glyde, Phys.~Rev.~A {\bf 68}, 033602 (2003).

\bibitem{16}
N.~Boninsegni, N.~V.~Prokofiev, and B.~V.~Svistunov,
Phys.~Rev.~E 74, 036701 (2006).

\bibitem{17}
S.~Pilati, K.~Sakkos, J.~Boronat, J.~Casulleras, and S.~Giorgini,
Phys.~Rev.~A 74, 043621 (2006).

\bibitem{18}
R.~A.~Duine and H.~T.~C.~Stoof, Phys.~Rep. {\bf 396}, 115 (2004).

\bibitem{19}
S.~L.~Cornish, N.~R.~Claussen, J.~L.~Roberts, E.~A.~Cornell, and C.~E.~Wieman, 
Phys. Rev. Lett. {\bf 85}, 1795 (2000).

\bibitem{20}
J.~L.~Roberts, J.~P.~Burke, N.~R.~Claussen, S.~L.~Cornish, E.~A.~Donley, 
and C.~E.~Wieman, Phys.~Rev.~A {\bf 64}, 024702 (2001).

\bibitem{21}
N.~R.~Claussen, E.~A.~Donley, S.~T.~Thompson, and C.~E.~Wieman,
Phys.~Rev.~Lett. {\bf 89}, 010401 (2002).

\bibitem{22}
M.~Olshanii, Phys.~Rev.~Lett. {\bf 81}, 938 (1998).

\bibitem{23}
D.~S.~Petrov and G.~V.~Shlyapnikov,
Phys.~Rev.~A {\bf 64}, 012706 (2001).

\bibitem{24}
E.~B.~Kolomeisky, T.~J.~Newman, J.~P.~Straley, and X.~Qi,
Phys.~Rev.~Lett. {\bf 85}, 1146 (2000).

\bibitem{25}
M.~D.~Lee, S.~A.~Morgan, M.~J.~Davis, and K.~Burnett,
Phys.~Rev.~A {\bf 65}, 043617 (2002).

\bibitem{26}
G.~S.~Astrakharchik, D.~Blume, S.~Giorgini, and B.~E.~Granger,
Phys.~Rev.~Lett. {\bf 92}, 030402 (2004).

\bibitem{27}
V.~I.~Yukalov, Phys.~Rep. {\bf 208}, 395 (1991).

\bibitem{28}
V.~I.~Yukalov, Phys.~Rev.~E {\bf 72}, 066119 (2005).

\bibitem{29}
V.~I.~Yukalov, Int.~J.~Mod.~Phys.~B {\bf 21}, 69 (2007).

\bibitem{30}
V.~I.~Yukalov,
Phys.~Lett.~A {\bf 359}, 712 (2006).

\bibitem{31}
V.~I.~Yukalov, Laser~Phys.~Lett. {\bf 3}, 406 (2006).

\bibitem{32}
V.~I.~Yukalov and H.~Kleinert,
Phys.~Rev.~A {\bf 73}, 063612 (2006).

\bibitem{33}
V.~I.~Yukalov and E.~P.~Yukalova,
Phys.~Rev.~A {\bf 74}, 063623 (2006).

\bibitem{34}
T.~D.~Lee and C.~N.~Yang,
Phys.~Rev. {\bf 105}, 1119 (1957).

\bibitem{35}
T.~D.~Lee, K.~Huang, and C.~N.~Yang,
Phys.~Rev. {\bf 106}, 1135 (1957).

\bibitem{36}
T.~D.~Lee and C.~N.~Yang,
Phys.~Rev. {\bf 112}, 1419 (1958).

\bibitem{37}
K.~Huang and H.~F.~Meng,
Phys.~Rev.~Lett. {\bf 69}, 644 (1992).

\bibitem{Lopatin}
A.~V.~Lopatin and V.~M.~Vinokur,
Phys.~Rev.~Lett. {\bf 88}, 235503 (2002).

\bibitem{Paul}
T.~Paul, P.~Leboeuf, N.~Pavloff, K.~Richter, and P.~Schlagheck,
Phys.~Rev.~A {\bf 72}, 063621 (2005).

\bibitem{Modugno}
M.~Modugno,
Phys.~Rev.~A {\bf 73}, 013606 (2006).

\bibitem{Bilas}
N.~Bilas and N.~Pavloff,
Eur.~Phys.~J.~D {\bf 40}, 387 (2006).

\bibitem{Sanchez-Palencia}
L.~Sanchez-Palencia,
Phys.~Rev.~A {\bf 74}, 053625 (2006).

\bibitem{Lugan}
P.~Lugan, D.~Cl\'ement, P.~Bouyer, A.~Aspect, M.~Lewenstein, and 
L.~Sanchez-Palencia, Phys.~Rev.~Lett. {\bf 98}, 170403 (2007).

\bibitem{Falco}
G.~M.~Falco, A.~Pelster, and R.~Graham,
Phys.~Rev.~A {\bf 75}, 063619 (2007);
{\it ibid}. {\bf 76}, 013624 (2007).

\bibitem{Lye}
J.~E.~Lye, L.~Fallani, M.~Modugno, D.~S.~Wiersma, C.~Fort, and M.~Inguscio,
Phys.~Rev.~Lett. {\bf 95}, 070401 (2005).

\bibitem{Clement}
D.~Cl\'ement, A.~F.~Var\'on, M.~Hugbart, J.~A.~Retter, P.~Bouyer, 
L.~Sanchez-Palencia, D.~M.~Gangardt, G.~V.~Shlyapnikov, and A.~Aspect,
Phys.~Rev.~Lett. {\bf 95}, 170409 (2005).

\bibitem{Fort}
C.~Fort, L.~Fallani, V.~Guarrera, J.~E.~Lye, M.~Modugno, D.~S.~Wiersma, 
and M.~Inguscio, Phys.~Rev.~Lett. {\bf 95}, 170410 (2005).

\bibitem{39}
J.~D.~Reppy, J.~Low~Temp.~Phys. {\bf 87}, 205 (1992).

\bibitem{40}
G.~S.~Astrakharchik, J.~Boronat, J.~Casulleras, and S.~Giorgini,
Phys.~Rev.~A {\bf 66}, 023603 (2002).

\bibitem{41}
P.~Navez, A.~Pelster, and R.~Graham,
Appl.~Phys.~B {\bf 86}, 395 (2007).

\bibitem{38}
V.~I.~Yukalov and R.~Graham,
Phys.~Rev.~A {\bf 75}, 023619 (2007).

\bibitem{42}
V.~I.~Yukalov,
Laser~Phys. {\bf 16}, 511 (2006).

\bibitem{43}
N.~N.~Bogolubov,
{\it Lectures on Quantum Statistics} (Gordon and Breach, New York, 1967), 
Vol. 1.

\bibitem{44}
N.~N.~Bogolubov,
{\it Lectures on Quantum Statistics} (Gordon and Breach, New York, 1970), 
Vol. 2.

\bibitem{45}
K.~Binder and A.~P.~Yang, Rev.~Mod.~Phys. {\bf 58}, 801 (1986).

\bibitem{46}
V.~I.~Yukalov and E.~P.~Yukalova, Phys.~Part.~Nucl. {\bf 31}, 561 (2000).

\bibitem{47}
V.~I.~Yukalov and E.~P.~Yukalova, Phys.~Part.~Nucl. {\bf 35}, 348 (2004).

\bibitem{48}
V.~I.~Yukalov, Phys.~Rev.~B {\bf 71}, 184432 (2005).

\bibitem{49}
S.~Gluzman, V.~I.~Yukalov, and D.~Sornette,
Phys.~Rev.~E {\bf 67}, 026109 (2003).

\bibitem{50}
V.~I.~Yukalov and E.~P.~Yukalova,
Eur.~Phys.~J.~B {\bf 55}, 93 (2007).

\bibitem{61}
V.~I.~Yukalov and E.~P.~Yukalova,
Phys.~Rev.~A {\bf 72}, 063611 (2005).

\end{thebibliography}
\end{document}